\documentclass[extra,mreferee]{gji}
\usepackage{timet}

\usepackage{amsmath}
\usepackage{amsfonts}
\usepackage{graphicx}

\title[Inversion with wavelet-based and scale-dependent regularization]
  {Inversion of electromagnetic induction data using a novel wavelet-based and scale-dependent regularization term}
\author[Wouter Deleersnyder]
  {Wouter Deleersnyder$^{1,2}$, Benjamin Maveau$^{1}$, Thomas Hermans$^{2}$, David Dudal$^{1,3}$\\
  $^{1}$KU Leuven Campus Kortrijk - KULAK, Department of Physics, Etienne Sabbelaan 53, 8500 Kortrijk, Belgium \\
  $^{2}$Ghent University, Department of Geology, Krijgslaan 281 - S8, 9000 Gent, Belgium \\
  $^{3}$Ghent University, Department of Physics and Astronomy, Ghent, Krijgslaan 281 - S9, 9000 Gent, Belgium \\
  }
\date{Received 2020 XXXX XX; in original form 2020 XXXX XX}
\pagerange{\pageref{firstpage}--\pageref{lastpage}}
\volume{XXX}
\pubyear{2020}

\newtheorem{theorem}{Theorem}[section]

\begin{document}

\label{firstpage}

\maketitle

\begin{summary}
The inversion of electromagnetic induction data to a conductivity profile is an ill-posed problem. Regularization improves the stability of the inversion and, based on Occam's razor principle, a smoothing constraint is typically used. However, the conductivity profiles are not always expected to be smooth. Here, we develop a new inversion scheme in which we transform the model to the wavelet space and impose a sparsity constraint. This sparsity constrained inversion scheme will minimize an objective function with a least-squares data misfit and a sparsity measure of the model in the wavelet domain. A model in the wavelet domain has both temporal as spatial resolution, and penalizing small-scale coefficients effectively reduces the complexity of the model. Depending on the expected conductivity profile, an optimal wavelet basis function can be chosen. The scheme is thus adaptive. Finally, we apply this new scheme on a frequency domain electromagnetic sounding (FDEM) dataset, but the scheme could equally apply to any other 1D geophysical method.
\end{summary}

\begin{keywords}
Inverse theory -- Wavelet transform -- Electromagnetic theory -- Controlled source electromagnetics (CSEM)
\end{keywords}

\section{Introduction}
Electromagnetic induction (EMI) surveys aim to image the electrical properties of the subsurface non-invasively. Via Archie's empirical petrophysical law \citep{archie1942electrical}, the electrical properties are related to the soil characteristics. Archie's law relates the water conductivity ($\sigma_w$), the porosity ($\theta$), the saturation of the pores ($S$) and experimental coefficients $m$ and $n$ (they are geometric factors) to the conductivity of the soil ($\sigma$):
\[
\sigma = \sigma_w \theta^mS^n.
\]
EMI methods are used in archaeological prospection \citep{saey2012electrical}, soil contamination \citep{pettersson2003environmental}, detection of electric anomalies (such as unexploded ordnance \citep{fernandez2010realistic}), agriculture \citep{jadoon2015application} and saltwater intrusion \citep{paepen2020combining}.

EMI instrumentation can often be used with different intercoil spacings or heights, resulting in data with a sensitivity varying with depth, allowing to invert the data and obtain a conductivity profile of the subsurface. Geophysical inversion is an ill-posed problem, because the solution is usually non-unique, which is generally cured with regularization techniques. Additionally, a slightly different dataset will generally yield an entirely different result. Deterministic methods impose one or two constraints, such as the smoothness constraint imposed by the traditional Tikhonov regularization \citep{tikhonov1943stability}. The smoothness constraint is not always adapted to the subsurface structure \citep{linde2015geological}. While many alternatives exist, it is not always easy to find the best adapted scheme. In earlier work on EMI inversion, Tikhonov regularization, truncated singular value decomposition \citep{deidda2014regularized} or sparsity based inversion \citep{tantum2012target} have beencommonly applied. In our work, a new regularization scheme is developed that is more adaptive than many other schemes as it can both represent smooth and sharp models. The regularization term is a sparsity constraint of the model parameters in wavelet domain. The wavelet basis determines the type of constraint (blocky, smooth or in between) that is imposed and hence this is an adaptive inversion procedure. \\

The literature about full-waveform inversion is richer and faces similar problems. Full-waveform inversion is concerned with fitting the model parameters, velocities and impedances, to the recorded seismic or ground penetrating radar data. The ill-posedness of the problem similarly requires a suitable regularization scheme in order to obtain a realistic model. The non-uniqueness is tackled typically with Tikhonov regularization \citep{virieux2009overview}. More recently, total variation regularization is applied to this field, including blocky regularization schemes with several focusing functions \citep{guitton2012blocky}, such as the $\ell_1$-norm.\\
The high non-linearity of full-waveform inversion results typically in a multimodal objective function, for which multi-scale approaches tend to be succesful \citep{bunks1995multiscale}. In EM data inversion, the objective function may also be multimodal, depending on the choice of the forward model operator. In such cases, \citep{bunks1995multiscale} suggested the use of the wavelet decomposition as a direction of future work, because it can decompose the problem by scale in an efficient fashion.\\

Wavelets are commonly used in signal processing (e.g. denoising), compressed sensing and image compression. For example, the JPEG 2000 standard uses the discrete wavelet transform algorithms and a specific type of wavelet particularly well-suited for this purpose: the Cohen-Daubechies-Feaveau wavelet. Wavelets have already been used in an inversion scheme for FDEM data \citep{liu2017wavelet}, where it has been shown to produce as good resolution as $\ell_1$-norm in the space domain. In their methodology, a finite difference method was used to solve the forward problem. Due to the slow forward modelling, only about ten iterations could be considered in the inverse problem. In this study, we develop a different inversion scheme, that uses a scale-dependent penalizing term (a comparison is provided in Section \ref{sec:scale_nonscale}). Inspired by the technicalities of the wavelet transform (cfr. Theorem \ref{theorem} in Section \ref{sec:wavbasis}) and the pragmatic idea that small scale details should be penalized more heavily than the coarser main structure of the conductivity profile, each length scale of the conductivity profile is penalized with a different weight. Furthermore, our scheme uses a faster (approximate) forward modelling operator (see Section \ref{sec:forward}). This allows us to compute much more iterations (in fact, computational power is not an issue anymore), at the cost of an additional modelling error due to the approximations in the forward operator. In Section \ref{sec:role_F}, we demonstrate that this modelling error poses not much of an issue. Our results show mainly two advantages. First, the scheme is more adaptive as it can recover both blocky and smooth conductivity profiles. Secondly, it can recover high amplitude anomalies in a globally smooth conductivity profile.\\

In this work, we first present the methods and the technicalities of the inversion scheme, including forward modelling, the inversion procedure and the basics of wavelet theory. In particular, we review the properties of wavelet basis and discuss why they are good candidates as a family of regularization functionals for geophysical inversion. In Sections \ref{sec:scale_nonscale}-\ref{sec:role_F}, we apply our regularization scheme on synthetic data and compare it with the commonly employed Tikhonov regularization as a benchmark. Finally, we apply our scheme to a real dataset of the Gontrode forest, Belgium, a context at which an 1D inversion scheme is realistic \citep{bobe2020efficient}.

\section{Methods}
\subsection{Forward model}
\label{sec:forward}
The forward model describes the soil's response $\mitbf{d} \in \mathbb{C}^{n_d\times 1}$ to the magnetic dipole placed at height $h_0$ above the surface and a soil with specific parameter distribution $\mitbf{m} \in \mathbb{R}^{n_m \times 1}$, measured at $n_d$ intercoil spacings $s$. The parameter distribution $\mitbf{m}$ is referred to as model or model parameters and contains the electrical conductivities in Siemens per meter for a specific (usually equidistant) parametrization of the subsurface. Finding the operator $\mathcal{F}$, such that
\begin{equation}
\mitbf{d} = \mathcal{F}(\mitbf{m})
\end{equation}
holds, is called the \emph{forward problem}.
\citet{wait1951magnetic} derived such a model, taking into account the loop-loop couplings between eddy currents and the dampening of the electromagnetic fields, for the 1D approximation (horizontally stratified earth) in the quasi-stationary Maxwell regime. The model involves recursive relations with integration over Bessel functions, which have to be executed numerically. This computationally demanding task slows down an inversion procedure in which the forward problem needs to be calculated at every iteration. Wait later derived a linear model \citep{wait1962note}, which is extensively used in the work of McNeill \citep{mcneill1980electromagnetic} and many others (e.g. \citep{simpson2009evaluating} or \citep{desmedt2011reconstructing} ), known as the LIN approximation. The LIN approximation is linear and neglects the couplings and electromagnetic dampening. It is only valid at Low Induction Numbers\footnote{The= induction number $\displaystyle B$ is defined as the intercoil distance $s$ divided by the skin depth $\delta$, i.e. $ B = \frac{\omega \mu_0 \sigma s^2}{2}$, where $\omega$ is the angular frequency of the dipole, $\mu_0$ is the magnetic permeability of the vacuum and $\sigma$ the electric conductivity of the medium. The meaning of sufficiently low induction numbers is that if the skin depth is much larger than the path the electromagnetic field has to traverse, the dampening can be neglected. Indeed, the path length is restricted by the fall-off of the magnetic dipole, which has the same order as the intercoil distance $s$.}. A more recently proposed model by \citep{maveau2020damped}, based on the LIN approach, takes into account the electromagnetic dampening and provides accurate results under more lenient conditions, especially in more conductive (e.g. high salinity) backgrounds. Using an approximate forward operator $\mathcal{F}_\mathrm{approx}$ introduces a biased modelling error $\mitbf{\kappa}$, we write
\begin{equation}
\mathcal{F}(\mitbf{m}) = \mathcal{F}_{\mathrm{approx}}(\mitbf{m}) + \mitbf{\kappa}.
\end{equation}

Another method to solve the forward problem is to use finite elements methods (FEM) or finite volume methods (FVM), such as provided for example in SimPEG \citep{cockett2015simpeg}. This type of modelling is considered exact if the mesh is sufficiently fine. For cylindrically symmetric set-ups (i.e. horizontally stratified conductivity profile \emph{and} a vertical dipole), the computation is relatively fast. In other cases, 
the computational burden of a detailed mesh is large. By consequence, only a few iterations in the inversion procedure can be considered.\\

The forward operator plays an important role in the inversion procedure (it will, for example, affect the number of local minima in the objective function). In this text, we focus on the inversion with wavelets and do not investigate in details the potential multimodality of the objective function. At first, we generate synthetic data and invert that data via the LIN approximation for its computational benefits. Furthermore, this method is sufficient to illustrate the working principle and the proof of concept of the inversion scheme with wavelets. In section \ref{sec:role_F}, however, we do elaborate on the role of the modelling error $\mitbf{\kappa}$ and the forward operator $\mathcal{F}$.

\subsection{Inversion approach}
\label{sec:inversion_approach}
Due to the non-linearity of many forward operators, the inverse problem is typically solved as an optimization problem. A suitable parameter distribution of the subsurface is determined by a minimum of an objective function $\phi$. The objective function takes the parameters distribution or model parameters as input and expresses how well the parameters fit the data. In minimum structure inversion procedures, an additional measure of model complexity, a regularization or model misfit term, is added to the objective function. The minimum is usually obtained via gradient descent-like methods.\\

The objective function in minimum structure inversion is a combination of a data misfit functional $\phi_d$, a model misfit functional $\phi_m$ and a regularization parameter $\lambda$. The latter balances the relative importance of both misfit terms. The data misfit functional $\phi_d$ measures how well the model parameters $\mitbf{m}$ fit the data $\mitbf{d}$ and is defined as a least-squares term
\begin{equation}
\label{eq:datamisfit}
\phi_d = \frac{1}{2}||D_d\left(\mitbf{d} - \mathcal{F}\left( \mitbf{m} \right)\right)||^2_2,
\end{equation}
where $D_d$ is a diagonal matrix whose elements are usually the reciprocals of the estimated noise standard deviation \citep{strutz2010data}. This weighting matrix penalizes the data misfit for data points with minimal measurement error less than data points with larger measurement error. The model misfit functional is a regularization term that is introduced to handle the ill-posedness of the problem. Indeed, geophysical inversion problems are ill-conditioned and the solution is typically not unique. A traditional method for minimum structure inversion is to apply a smoothing constraint, such as with Tikhonov regularization \citep{tikhonov1943stability}. Tikhonov regularization improves the stability of the inversion, but promotes smooth solutions and hence cannot recover blocky structures. Other regularization terms exist, such as the $\ell_1$-norm inversion in original model space \citep{farquharson2007constructing}, the  minimum gradient support functional \citep{portniaguine1999focusing} or a sequential inversion \citep{guillemoteau20161d}.\\

In this paper, we develop a different approach. Suppose that there exists a basis in which the true model parameters $\mitbf{m}$, known to possess minimum structure, are represented in a sparse fashion in $\mitbf{x} \in \mathbb{R}^{n_x \times 1}$. Then,
\begin{equation}
\mitbf{x} = \mitbf{W}\mitbf{m},
\end{equation} where $\mitbf{W}$ is the basis transformation. Such a basis transformation, in combination with a sparsity promoting measure, yields an appropriate model misfit, because a complex structure in the model parameters $\mitbf{m}$ will have many non-zero entries in the $\mitbf{x}$-representation and will therefore be heavily penalized by the sparsity promoting measure. The  $\ell_1$-norm is a well-known sparsity promoting measure. We will rely on gradient decent-like methods and since the $\ell_1$-norm is not differentiable at zero, we will use the perturbed $\ell_1$-norm measure of Ekblom \citep{ekblom1987l1}
\begin{equation}
\mu_{\mathrm{Ekblom}}(x) = \sqrt{ x^2 + \epsilon},
\end{equation}
which is very similar to the $\ell_1$-norm for small $\epsilon$ and which is also convex. The latter is a welcome property for the optimization problem. The model misfit in terms of $\mitbf{x}$ in our inversion scheme is thus
\begin{equation}
\phi_m(\mitbf{x}) = \sum_j \sqrt{x_j^2 + \epsilon}.
\end{equation}

In summary, we have
\begin{eqnarray}
\label{eq:objectivefunction1}
\min_{\mitbf{x}} \phi &=& \min_{\mitbf{x}} \left( \phi_d + \lambda\phi_m\right)\\ 
& =& \min_{\mitbf{x}} \left( \frac{1}{2}||D_d(\mitbf{d} - \mathcal{F}_\mathrm{approx}\left( \mitbf{W^{-1}x} \right)||^2_2 \right.\nonumber \\
& & \hspace{10em} + \left. \lambda  \sum_{j = 1}^{n_x} \sqrt{x_j^2 + \epsilon} \right),
\end{eqnarray}
where the regularization parameter $\lambda$ is still undetermined. A regularization parameter can be too small, that is the case when the data misfit $\phi_d$ is smaller than the actual noise on the data, i.e. when we have overfitting. On the other hand, with a too large regularization parameter, the parameter distribution will exhibit too little structure. \citet{hansen2010discrete} discusses the common automatic methods to estimate an optimal regularization parameter. We will rely on the L-curve criterion. This method requires us to solve the inverse problem for multiple regularization parameters and plots the model misfit in terms of the data misfit. It is expected that the outcomes of each inversion will appear in an L-shape: there will be branches in which a change in the regularization parameter will yield a significant change in either the data or model misfit. There will be a corner in which the change in the regularization parameter will yield a roughly equal significant change in both misfits. The corner is usually the point in the $(\phi_d, \phi_m)$-plane closest to the origin and thus with both misfits minimized. The corner is usually more pronounced and easier to detect in log-log plane. In general, there is no guarantee that the curve will exhibit an L-shape. Local corners pose difficulties for the automatic corner detection. We rely on the adaptive pruning algorithm \citep{hansen2007adaptive}, because it is a robust method for the appropriate corner selection.
\subsection{Wavelet basis}
\label{sec:wavbasis}
The basis transformation $\mitbf{W}$ transforms a minimum structure model $\mitbf{m}$ to a sparse representation $\mitbf{x}$ in the wavelet domain. The wavelet transform has both spatial and temporal resolution and allows to represent the conductivity profile in a sparse fashion. By studying the wavelet transform, we can analyse both the location of the peaks (spatial resolution) and periodicities at different frequencies (temporal resolution). The sparsity of the transform is guaranteed by a theorem presented later in this text. Most of the material in this section can be found in \citep{strang1996wavelets} and \citep{mallat1989theory}.\\

While the wavelet transform has a complete and independent theory based on the definition of multiresolution analysis \citep{daubechies1992ten}, we sketch a more intuitive summary with analogies to Fourier series. 
In Fourier series, a $2\pi$ periodic signal in the continuous time domain is represented in terms of basis functions $\{e^{ikt}\}_{k\in \mathbb{Z}}$ with Fourier coefficients $f_k = \langle e^{ikt}, f  \rangle$ \citep{mallat1999wavelet}. Indeed, we write 
\begin{equation}
f(t) = \sum_{k\in \mathbb{Z}} f_k e^{ikt}. 
\label{wav:eq:fourierseries}
\end{equation}
Wavelet theory constructs a set of basis functions, consisting of \emph{scaling functions} $\varphi$ and \emph{wavelet functions} $\psi$. Analogously to Fourier series, the signal is expanded in a set of basis functions. Unlike Fourier series, these basis functions have compact support (i.e. they only exist on a finite interval) and $f$ does not need to be periodic. Finding such bases and determining their properties are the major matter of interest of wavelet theory. Instead of considering sines and cosines at different frequencies, the following translations and dilations of the wavelet and scaling functions span the function space:
\begin{equation}
\label{wav:eq:dilatation}
\varphi_{n,k}(t) = 2^{n/2}\varphi(2^nt-k) \quad \mathrm{or} \quad  \psi_{n,k}(t) = 2^{n/2}\psi(2^nt-k),
\end{equation}
where $n \in \mathbb{Z}$ is the dilation parameter that makes the basis function's compact support larger or smaller (temporal resolution). The $k$-parameter describes translations along the $t$-axis (spatial resolution). We refer to $\psi_n$ or $\varphi_n$ as a set of basis functions at resolution level $n$ and these sets span the spaces $W_n$ and $V_n$ respectively. Assume that the bases are orthonormal (this is not required, biorthogonal wavelets can also be used). The scaling function is more fundamental than the wavelet function, the wavelet function at resolution level $n$ is a linear combination of scaling functions at resolution level $n+1$.\\

As with Fourier series, we can decompose an arbitrary function $f(t) \in (W_n \oplus V_n)$ as a series in both scaling and wavelet functions:
\begin{equation}
f(t) = \sum_{k\in \mathbb{Z}}v_{nk} \varphi_{nk}(t), \quad \quad f(t) = \sum_{k\in \mathbb{Z}}w_{nk} \psi_{nk}(t),
\end{equation}
where  $v_{nk} = \langle \phi_{nk}, f  \rangle$ and $w_{nk} = \langle \psi_{nk}, f  \rangle$  are scaling and wavelet coefficients respectively. In the discrete wavelet transform, the function $f(t)$ and wavelets are discretely sampled. We drop the time-dependence in this notation and we refer to it as a signal instead of a function.\\

A one level (discrete) \emph{wavelet transform} is the decomposition of the signal $f_n \in V_n$ in its components $f_{n-1}\in V_{n-1}$ and $g_{n-1}\in W_{n-1}$. 
\begin{equation}
f_n = \sum_k v_{nk}\varphi_{nk} =\underbrace{ \sum_k v_{n-1,k}\varphi_{n-1,k} }_{f_{n-1}}+ \underbrace{\sum_k w_{n-1,k}\psi_{n-1,k} }_{g_{n-1}}
\end{equation}
where $v_{n-1,k}$ are the scaling coefficients and $w_{n-1,k}$ the wavelet coefficients at resolution level $n-1$. The signal $g_{n-1}$ describes the details that were present in the signal on scale $n$, but have disappeared from the coarser $(n-1)$-scale $f_{n-1}$. A higher level wavelet transform is obtained by repeating the one level wavelet transform on $f_{n-1}$. This process can be interpreted in the framework of filter banks, where the wavelet coefficients are obtained after passing through a high pass filter (contains details of signal) and a subsequent downsampling procedure. A low pass filter and the subsequent downsampling generate the scaling coefficients. These scaling functions are then again passed through the high and low pass filter etc. The high and low pass filter coefficients are a result from specific conditions, for example that they allow for perfect reconstruction to the model space, i.e. such that there exists an inverse procedure to go from the coefficients in wavelet space to the original signal. For every wavelet basis function, there exists a unique set of low pass and high pass filters. Their relation is not obvious, for which we refer to \citet{mallat1999wavelet}. However, the filter bank interpretation allows for a fast and reliable computation of the wavelet transform with complexity $\mathcal{O}(N)$, in which the explicit form of the basis function is not required. This computation, known as the Fast Wavelet Transform (FWT), is analogous to the Fast Fourier Transform.\\

In our inversion approach, we require an explicit matrix $\mitbf{W} \in \mathbb{R}^{n_x \times n_m}$ for the basis transformation. This is obtained by computing $n_m$ wavelet transforms of a Dirac train\footnote{A Dirac train $\delta_i$ is a vector with zeros, and with a spike (a one) at the $i$-th index.} 
\begin{equation}
\mitbf{W} = \mitbf{W}\mathbb{I} = W[\delta_1 \delta_2 \cdots \delta_{n_m} ] =[\mitbf{W}\delta_1 \mitbf{W}\delta_2 \cdots \mitbf{W}\delta_{n_m} ],
\end{equation}
where $\mitbf{W}\delta_i$ is computed with an available implementation of the FWT, such as the PyWavelets package \citep{lee2006pywavelets} in Python. We distinguish the lengths of the signal in the different domains, more specifically $n_x \geq n_m$. This results from signal extension, which is an analogous situation as with convolutions\footnote{In the filter bank interpretation, the relation between $v_{nk}$ and $v_{n+1,k}$ is a convolution with a downsampled and time reversed impulse response.}. With finite signals, one must think about how to handle the boundaries. It is impossible to correlate a basis function with a part of a signal when that basis function has wider support than the length of that part of the signal. The signal is therefore by default symmetrically or periodically extended (e.g. in Matlab's implementation \citep{misiti2009matlab}). Another strategy to limit the boundary distortions, is to compute a maximum \emph{useful} level of decomposition. One often considers the maximum level $N_\mathrm{max}$
\begin{equation}
N_\mathrm{max} = \left\lfloor \log_2 \left( \frac{n_m}{F - 1} \right) \right\rfloor,
\end{equation} where $\lfloor \cdot \rfloor$ is the floor-operator and $F$ is the number of filter coefficients, which scales with the support of the wavelet.\\

Finally, some properties about wavelets are introduced to better understand the approximating abilities of a specific wavelet. 
The number of vanishing moments is the most decisive property of a wavelet. A wavelet has $p$ vanishing moments when
\begin{equation}
\int t^k\psi dt = 0 \quad \mbox{for $k = 0,1\cdots,p-1$}.
\label{eq:vanishingmoments}
\end{equation}
The number of vanishing moments is related with the compact support of the wavelet: an orthonormal wavelet with $p$ vanishing moments has at least a support of size $2p-1$ \citep{daubechies1988orthonormal}. Wavelets with minimal compact support for a given $p$ are called Daubechies wavelets (db$p$). Further, if $f(t)$ is $p$ times differentiable, its wavelet coefficients decay \citep{strang1996wavelets}:
\begin{theorem}{Decay of the Wavelet: }
	\label{theorem}
	If $f(t)$ is $p$ times differentiable, its wavelet coefficients decay like $2^{-jp}$:
	\begin{equation}
	|w_{j,k}| \leq C2^{-jp} ||f^{(p)}(t)||.
	\end{equation}
\end{theorem}
Hence, wavelet coefficients decay with increasing resolution level. Inversely, discontinuities or singularities in the signal $f$ yield large wavelet coefficients corresponding to the same spatial location on all resolution levels. This theorem, therefore, reveals an important trade-off: higher vanishing moments improve the approximating abilities of a wavelet (which is interesting in applications as it yields sparser representations) but involves wavelets with larger compact support. The larger the compact support, the more difficult to meet the conditions of the theorem: the signal $f$ must then be piecewise smooth on a larger interval. Daubechies wavelets are by definition the wavelets with best approximating abilities: they are for a given number of vanishing moments, the orthogonal wavelet with the smallest compact support.  \\

\begin{figure*}
	\centering
	\includegraphics[width=\textwidth]{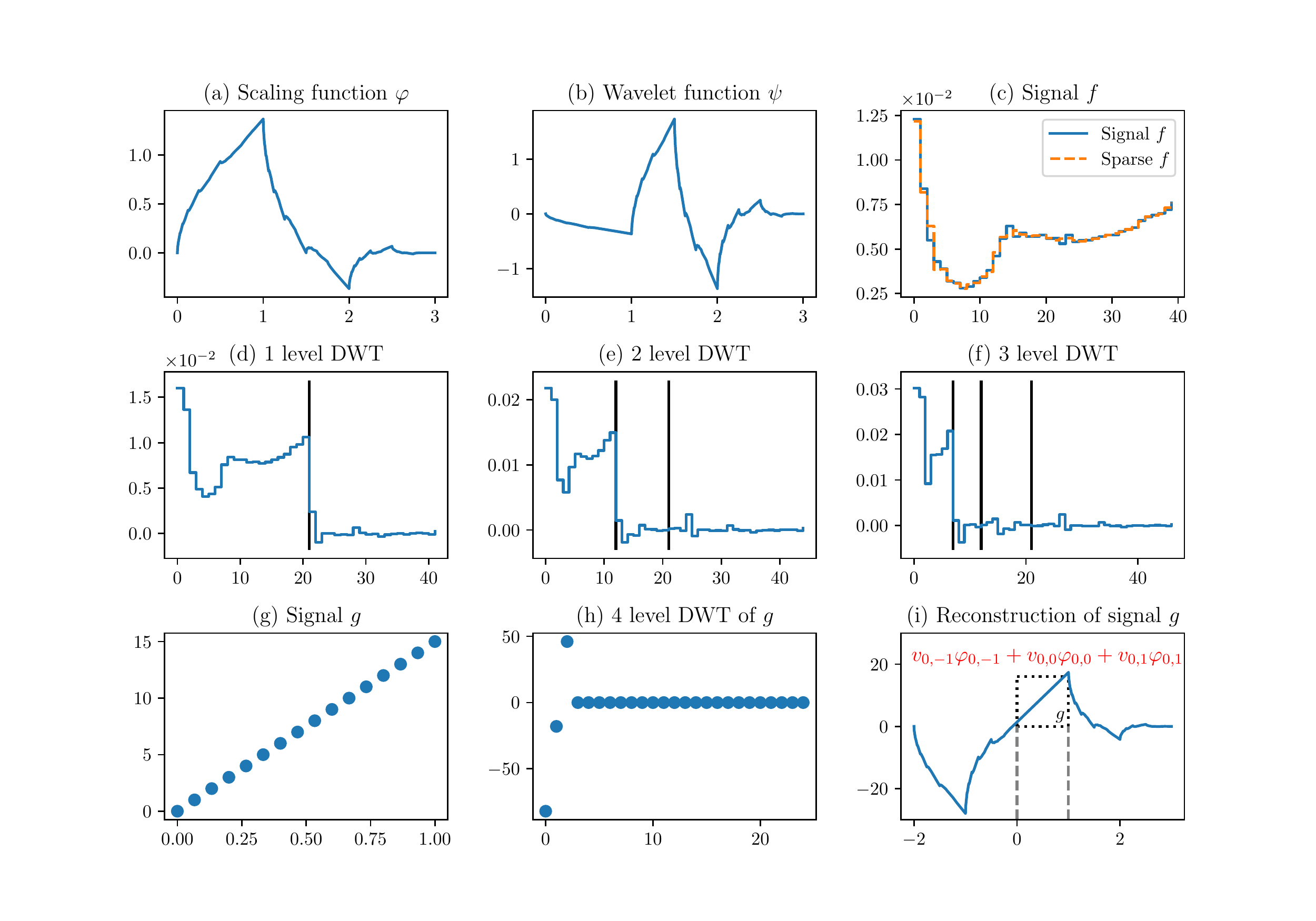}
	\caption{Example of the wavelet transform with Daubechies wavelets with two vanishing moments (db2) on two discrete signals $f$ and $g$. Vertical lines in figures (d)-(f) show the boundary between scaling coefficients and wavelet coefficients at different resolution levels.}
	\label{fig:exampledb2}
\end{figure*}

\begin{figure*}
	\centering
	\includegraphics[width=0.7\textwidth]{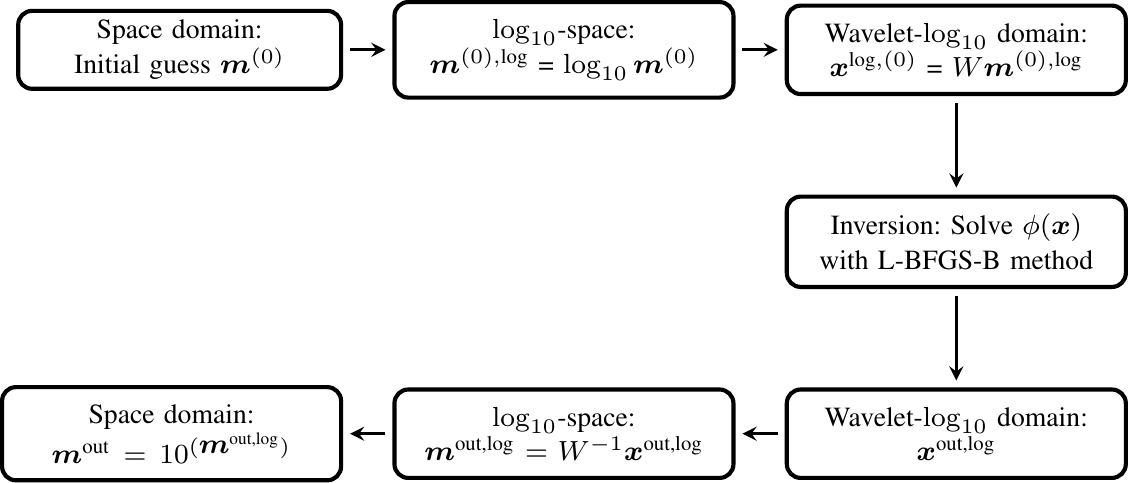}
	\caption{The inversion scheme in $\log_{10}$-space domain}
	\label{res:fig:logscheme}
\end{figure*}

The Daubechies wavelet with two vanishing moments (db2) is shown in Fig \ref{fig:exampledb2}. Their scaling and wavelet functions in \ref{fig:exampledb2}(a) and \ref{fig:exampledb2}(b) have a support width of $2p-1$ or 3. Figs \ref{fig:exampledb2}(d)-(f) show respectively the one, two and three level discrete wavelet transform of signal $f$ in Fig \ref{fig:exampledb2}(c). Signal $f$ is a discretely sampled conductivity profile obtained from \citep{hermans2017facies}. It exhibits minimum-structure in the sense that neighbouring samples have similar electrical conductivities. The samples can be viewed as scaling coefficients $v_n$ at resolution level $n$. The one level DWT (with by default symmetric signal extension) transforms these scaling coefficients to scaling coefficients $v_{n-1}$ and wavelet coefficients $w_{n-1}$ at a coarser resolution level $n-1$. The former describes the signal at a coarser scale, while the latter coefficients contain the details. In Fig \ref{fig:exampledb2}(d), the scaling coefficients, the 20 coefficients in the first half of the figure, describe the original signal with less detail (and are also rescaled with a factor $\sqrt{2}$, see Equation \eqref{wav:eq:dilatation}). One can immediately notice that for such minimum-structure signals the wavelet coefficients, in the second half of the figure, are small. When the small wavelet coefficients are ignored (set to zero), one gets a sparser representation of the original signal. The back transformation to model space (sparse signal $f$ or dashed line in Fig \ref{fig:exampledb2}(c)) is still be very similar to the original signal $f$. This is obtained by setting all wavelet coefficients equal to zero if it is smaller than 5\% of the largest scaling coefficient. This common practice in wavelet compression yields a representation with all scaling coefficients and only one wavelet coefficient. To go from the one level to the two level DWT, the same procedure is applied on the scaling coefficients $v_{n-1}$ (the wavelet coefficients $w_{n-1}$ remain unaltered). The scaling coefficients $v_{n-2}$ again describe the original signal at an even coarser scale and $w_{n-2}$ describe the details that are lost since $v_{n-1}$. Fig \ref{fig:exampledb2}(f) shows the three level DWT. Note that the number of coefficients in wavelet space is slightly larger than in model space. This is due the signal extension at the boundaries. In general
\[
n_m \leq n_{x,1} \leq n_{x,2} \leq n_{x,3},
\]
where $n_m$ is the length of the signal $f$ in model space and $n_{x,i}$ the i-th level DWT of signal $f$. Wavelets with larger support widths yield wavelet representations with more coefficients.\\ A wavelet with $p$ vanishing moments is orthogonal to polynomials of degree $p-1$ (see Eq. \eqref{eq:vanishingmoments}). Hence, the db2 wavelet is orthogonal to linear functions. This guarantees that wavelet coefficients will be zero for linear pieces in the signal $f$. The greater the number of vanishing moments, the more complex structures can be represented in a sparse fashion. This clearly shows that wavelets with higher vanishing moments have better approximating abilities. In Figs     \ref{fig:exampledb2}(j)-(i) we consider a discretely sampled linear function $g$ and its maximum four level wavelet transform. One immediately notices that all wavelet coefficients (right-side) are zero. In the scaling coefficients, one can still recognize the original (rescaled) structure. The minimum-structure linear function can consequently be represented \emph{exactly} with only three coefficients in the wavelet representation. To increase intuition, we return to Equation \eqref{wav:eq:fourierseries} in which we expand a signal into a basis. It is not obvious by eye that a linear combination of the irregular shaped db2 scaling function yields the original function. In this example, there are only three terms: three scaling functions with support width 3. In Fig \ref{fig:exampledb2}(i), we show each scaling function multiplied by its scaling coefficient and in the correct location (depends on the translation parameter $ k $). If we add the signals, we get a linear function in the original interval $ [0,1] $. Note that we used smooth signal extension here, as a result of which we effectively avoided boundary distortions. \\

\subsection{Scale-dependent regularization}
Finally, we introduce what we call scale-dependent regularization. The problem with the regularization term in equation \eqref{eq:objectivefunction1} is that it gives each coefficients in $\mitbf{x}$ the same weight. However, as we \emph{know} that some components are more likely to be zero than others, we account for this in the objective function. The first coefficient will never be zero, because it would mean that the integrated value over the conductivity profile would be zero. Wavelet coefficients at smaller scales are expected to be zero, since these correspond with neighbouring model parameters having equal conductivities. Our wavelet basis, however, exhibits better minimum structure when the high resolution wavelet coefficients $x_i \in W_{-1}$ are sparse. Theorem \ref{theorem} makes this more formal for any type of wavelet with $p$ vanishing moments, recall that if $f$ is locally smooth and $p$ times differentiable, the theorem states that at scale $j + 1$ the wavelet coefficients, localized where $f$ is smooth and $p$ times differentiable, are approximately smaller than those on scale $j$ by a factor of $2^p$. This result can be used to define a new regularization term (where we assume no signal extension):
\begin{align}
\phi_m(\mitbf{x}) &=  \frac{1}{E} \left( \mu(x_2) + 2^{1}\sum_{i = 3}^{4}\mu(x_i) \right. \nonumber \\
& \hspace{4em} \left. + 2^{2}\sum_{i = 5}^{8}\mu(x_i)  + 2^{3}\sum_{i = 9}^{16}\mu(x_i)  + \cdots\right),
\end{align}
where we have made the sparsity constraint for high resolution coefficients more stringent and $\mu$ is the Ekblom measure. Note that we have dropped the number of vanishing moments $p$ in the exponential as in Theorem \ref{theorem}, because this would be an excessive penalization of the small-scale wavelet coefficients for a large number of vanishing moments $p$ (the results for $2^{j}$ are in general better than for $2^{jp}$). Furthermore, we normalize the data misfit functional with the Euclidean norm ($E$) of coefficients to be able to better compare between different parametrizations.

\subsection{Overview of the inversion scheme}
Solving the inverse problem is basically minimizing an objective function. Constructing a suitable function, however, requires some thought and multiple assumptions must be made. Our inversion scheme is illustrated in Fig \ref{res:fig:logscheme}. An iterative optimization method starts with an initial model or guess. In principle, this model can be based on prior knowledge of the geological context (e.g. groundtruth knowledge obtained in the vicinity of the surveying site) or by relying on the apparent conductivity, although the initial model can likewise be generated randomly. The latter option is less suitable for multimodal objective functions.\\

Following the generation of the initial model, it is transformed into the logarithmic domain. This is a common action in geophysical inversion schemes to ensure positive parameters, such as for the electrical conductivity. Of course, we will conduct a back-transform to the original domain later in the inversion scheme, where all values from the logarithmic domain are transformed to the $\mathbb{R}^+$ domain.\\

The model in the logarithmic domain is subsequently transformed into wavelet domain by the discrete wavelet transform. This involves a few preliminary choices that must be made before the transform can be applied. Each decision affects the objective function and hence potentially affects its minimum. First of all, one must decide which wavelet to use. Since in this paper only one dimension is considered, we choose wavelets from the Daubechies family for their approximating abilities, given the trade-off between compact support width and the number of vanishing moments. A second decision is the level of the wavelet transform. In principle, we require a sparse representation of a geologically realistic model, because we work with a sparsity promoting measure. We have already argued in the previous section that for a signal of length $2^N$, it is not necessarily better to apply an $N$ level DWT, due to boundary distortions. Nevertheless, we choose $N = \log_2(n_m)$ which turns out to work equally well. To minimize those boundary distortions, the choice which signal extension mode to use plays a role. In a geological context, we generally do not expect the model parameters to be periodic, so we opt for symmetrization or a smooth signal extension. Finally note that the shape of the wavelet will affect the result of the inversion, especially if we opt for a strong degree of regularization. Accordingly, if a blocky model is expected, it is better to opt for blocky wavelets which have sharper edges, such as the db1 wavelet. If soft boundaries are expected between the layers, then smoother wavelets are more appropriate. These wavelets usually have a larger number of vanishing moments and have a stabilizing effect due to their larger compact support.\\

There are various optimization methods available, classified into trust-region, line-search and non-gradient methods. In this paper we exclusively use the Broyden–Fletcher–Goldfarb–Shanno (BFGS) algorithm. This algorithm is implemented in Python (SciPy) \citep{scipy} with the line search algorithm that meets the Wolfe conditions \citep{more1994line}. This method is a Newton's method and approximates the Hessian at every iteration. In practice, the choice of tolerances (stop criteria) will affect the actual outcome. In general, the tolerance on the objective function is set to $10^{-11}$ and $10^{-10}$ for the gradient is sufficient (for further details, see \citep{scipy}).

\section{Results}
\subsection{Scale-dependent regularization}
\begin{figure*}
	\centering
	\includegraphics[width=0.48\textwidth]{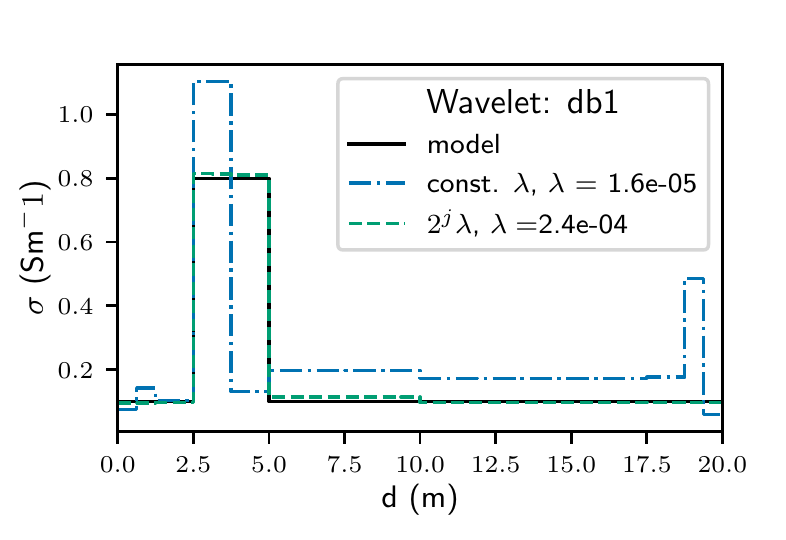}
	\includegraphics[width=0.48\textwidth]{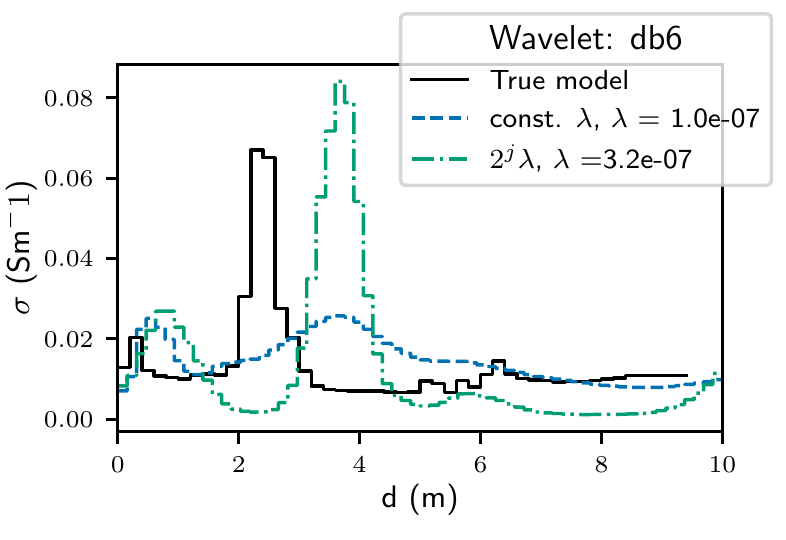}
	\caption{Comparison of traditional vs. scale-dependent wavelet-based regularization schemes with the LIN approximation as forward operator.}
	\label{fig:scale_nonscale}
\end{figure*}
\label{sec:scale_nonscale}
In the first examples, we demonstrate the advantage of the scale-dependence of our novel regularization scheme. Whereas other schemes (such as \citep{liu2017wavelet}) penalize each wavelet coefficient with equal weight, our scheme penalizes each wavelet coefficient proportionally with their resolution level $n$.\\

To make a valid comparison between the two wavelet-based regularization schemes, we first eliminate all other factors in the inversion process that affects the outcome. Such factors are, for example, the presence of numerous local minima in the objective function, the presence of a modelling error $\mitbf{\kappa}$ in the forward operator or a limited maximal number of iterations due to limited computational resources. We eliminate the modelling error by using the same forward operator for the synthetic data generation and the inversion. The objective function will have good properties with the LIN approximation because the model is linear and consequently the data misfit term will be quadratic, while the Ekblom measure is convex. Furthermore, choosing this forward model overcomes the computational burden. We elaborate on the effect of the forward model in Section \ref{sec:role_F}.\\

The first example in Fig \ref{fig:scale_nonscale} shows a realistic synthetic, three-layer conductivity profile, where a highly conductive layer of 0.8 S/m from a depth of 2.5 m to 5 m is embedded in a less conductive half-space of 0.1 S/m. This is a case where smooth inversion typically yields a very smeared out profile, while a blocky outcome is better suited. We show the result with db1 wavelets (see Fig \ref{fig:scaling_functions} in Appendix \ref{appendix:daubechies} for the shape of the wavelets, in the next section with other wavelets and smooth inversion) with both scale-dependent regularization ($ 2^j \lambda $) and the traditional sparsity inversion with equal regularization (const. $\lambda $). The L-curve criterion is used for the estimation of the optimal regularization parameter. While the convexity of the L-curve is not guaranteed, the L-curves displayed the typical L-shape (not shown).\\

The synthetic data is generated for a vertical dipole in both components (horizontal coplanar HCP and perpendicular PERP) for 20 equidistant intercoil distances from 1 to 20 meters at a height of 0.1 meters above the soil and where we have added 5\% Gaussian multiplicative noise. We choose to fit 32 model parameters. The result of the scale-dependent sparsity inversion outperforms the traditional wavelet-based inversion. The scale-dependent scheme follows the true model almost exactly, while the latter has too much structure at the small scales. This example illustrates the advantage of scale-dependent regularization over traditional wavelet-based inversion with db1 wavelets. \\

The second example shows a conductivity profile from a borehole logging from an alluvial aquifer \citep{hermans2017facies}. It has ten times smaller conductivity values and therefore the variation in the magnetic field data will be more subtle, especially compared with the noise level of 5\%. In this case, we opt for a more realistic undetermined problem in which we fit 64 model parameters with the db6 wavelet to data from a vertical dipole for both components and 5 equidistant intercoil distances from 1 to 10 meters. The results in Fig \ref{fig:scale_nonscale} show for both schemes a wrong location of the main peak at 2.5m depth. However, we observe that the scale-dependent regularization has a much narrower peak and whose amplitude is closer to the exact values indicating that the scale-dependent regularization seems also better suited for other wavelet bases. The misidentification of the peak is related to the combination of the noise level with the challenging nature of the inverse problem. The inversion of the noise-free data set (shown in the Appendix \ref{appendix:nonoise} in Fig \ref{fig:nonoise1}) yields a correct location of the peak (for both regularization methods).\\

\subsection{Model space vs Wavelet space}
\label{sec:modwav}
\begin{figure*}
	\centering
	\includegraphics[width=0.48\textwidth]{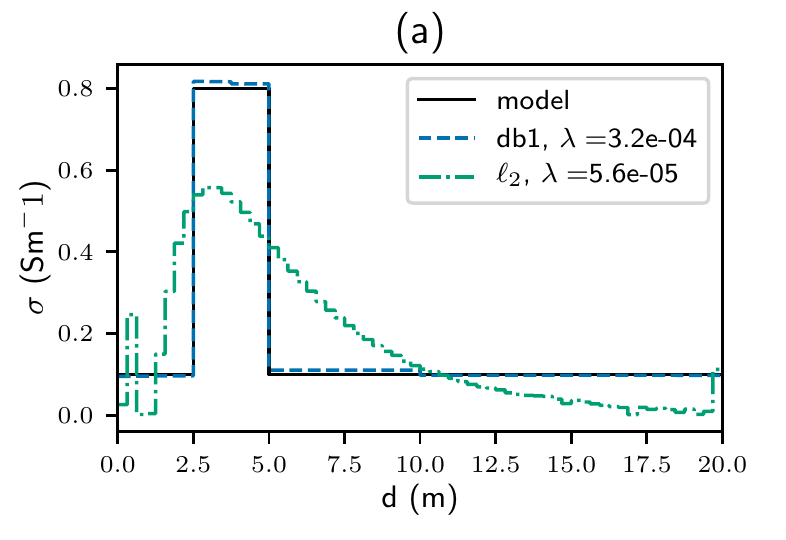}
	\includegraphics[width=0.48\textwidth]{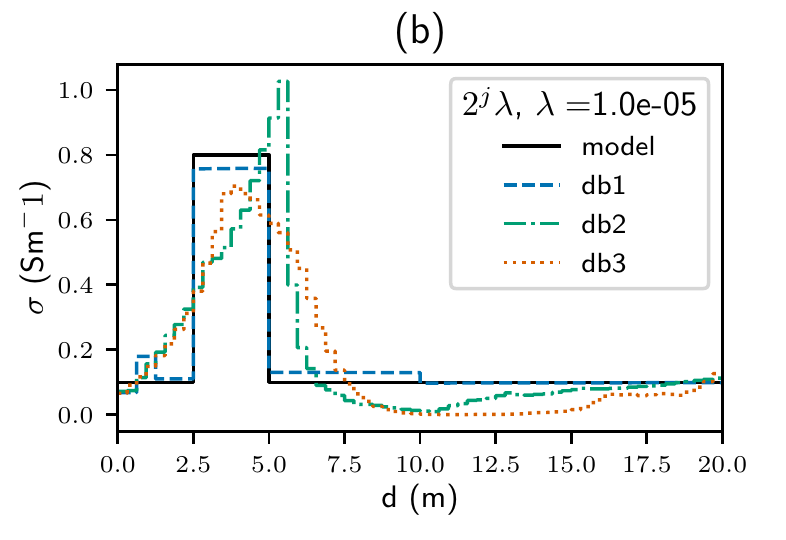}
	\includegraphics[width=0.48\textwidth]{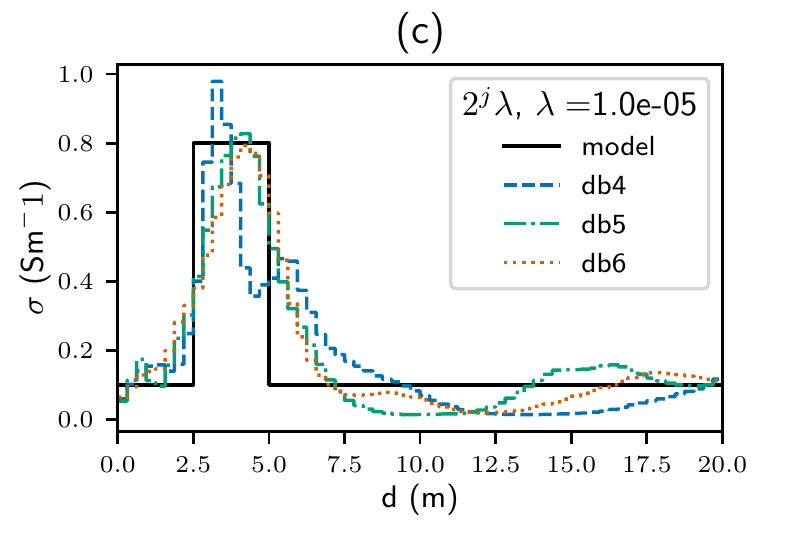}
	\includegraphics[width=0.48\textwidth]{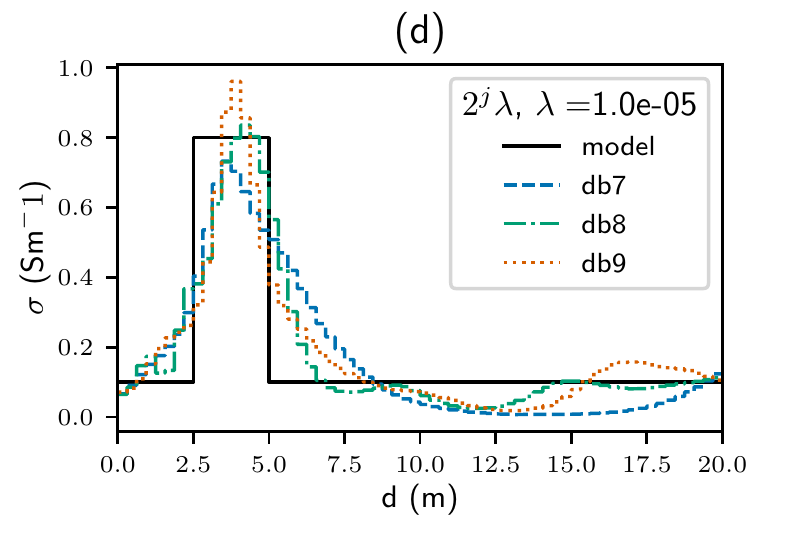}
	\caption{Models after inversion with no modelling error ($\mitbf{\kappa} = 0$) and the adaptive wavelet-based regularization with various Daubechies wavelets vs. Tikhonov regularization.}
	\label{fig:different_wavelets}
\end{figure*}

\begin{figure}
	\centering
	\includegraphics[width=0.48\textwidth]{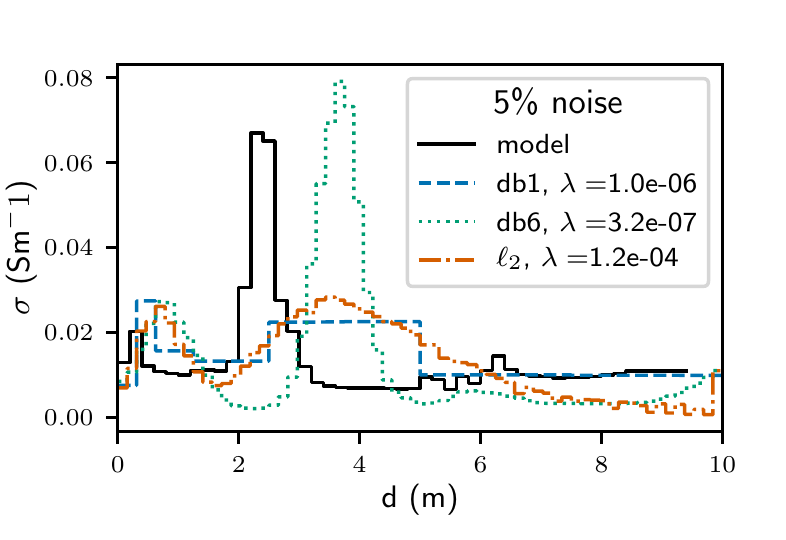}
	\caption{Models after inversion with no modelling error ($\mitbf{\kappa} = 0$) and the adaptive wavelet-based regularization with the db1 and db6-wavelet vs. Tikhonov regularization.}
	\label{fig:modelspacedb6}
\end{figure}

\begin{figure*}
	\centering
	\includegraphics[width=0.495\textwidth]{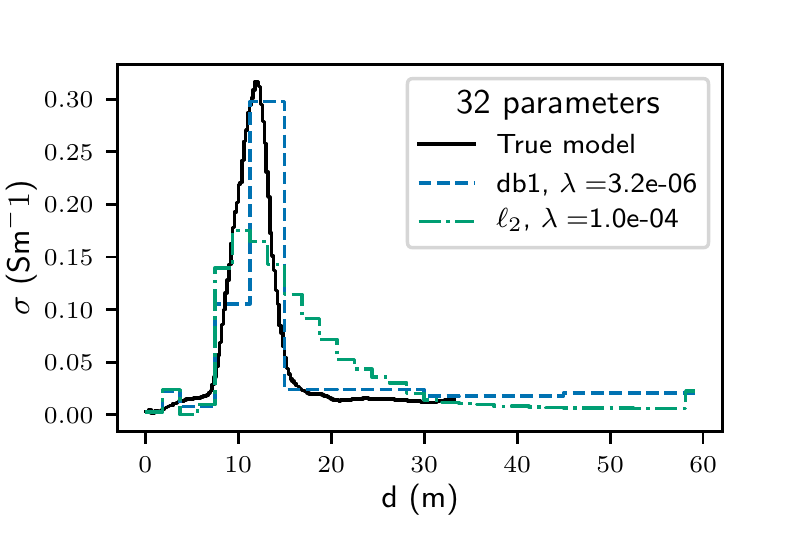}
	\includegraphics[width=0.495\textwidth]{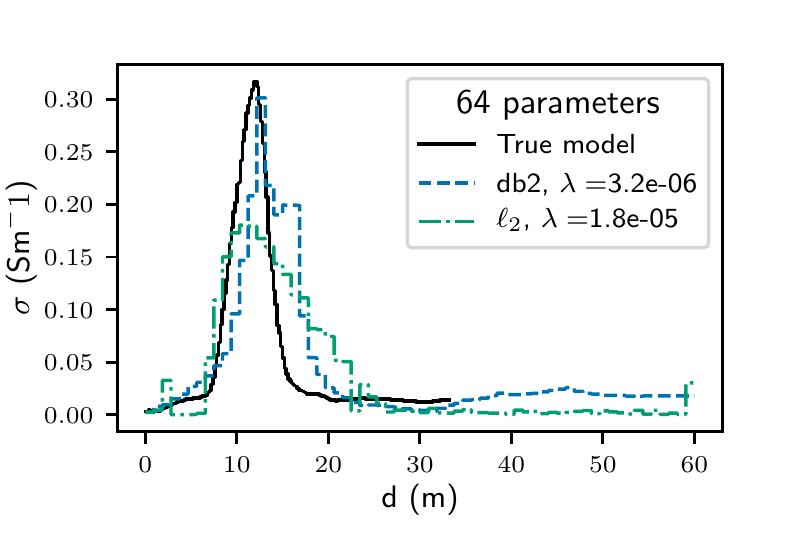}
	\includegraphics[width=0.495\textwidth]{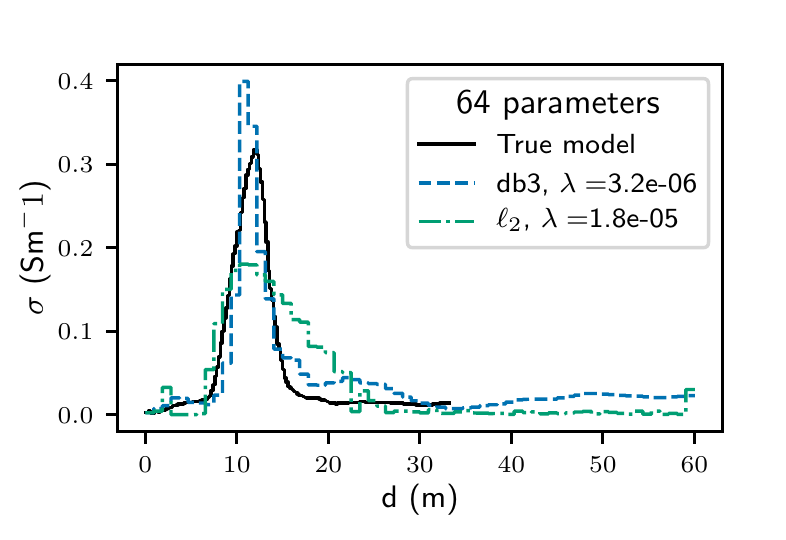}
	\includegraphics[width=0.495\textwidth]{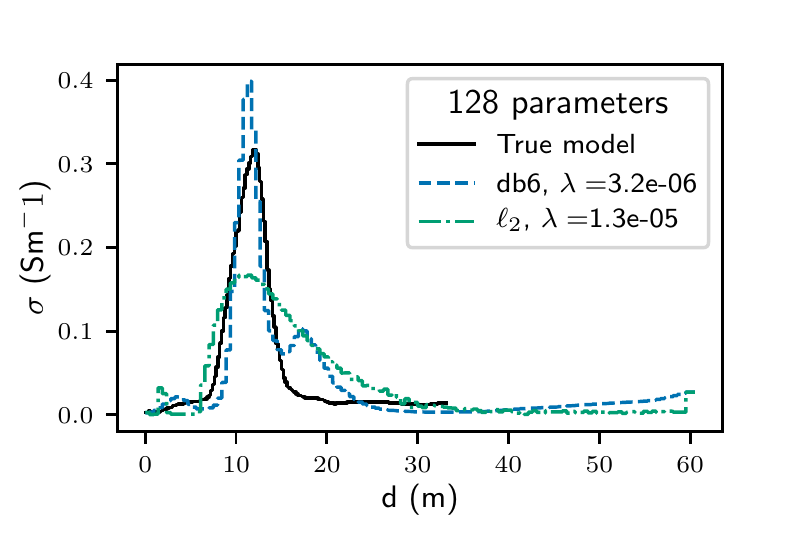}
	\caption{Models after inversion with no modelling error ($\mitbf{\kappa} = 0$) and the adaptive wavelet-based regularization with various Daubechies vs. Tikhonov regularization.}
	\label{fig:westhoek}
\end{figure*}
In this section, we illustrate our proposed regularization scheme and take the traditional scheme with smoothing Tikhonov regularization as benchmark. Our adaptive scheme can recover blocky structures and sharp boundaries, which is lacking in minimum-structure inversions with smoothing constraints. To illustrate the advantage of wavelet space over standard model space, we perform a numerical test on the three-layered model described in the previous section, with the only difference that we fit 64 model parameters instead of 32. The results after inversion are shown in Fig \ref{fig:different_wavelets}. From the shapes of the wavelet, it is expected that the db1-wavelet yields the best (blocky) inversion. We therefore compare the results of the db1-wavelet with smoothing inversion in Fig \ref{fig:different_wavelets}(a). The wavelet-based inversion follows the model closely, while the smoothing inversion locates the peak correctly, but with an underestimated value for the electrical conductivity as expected. The smooth inversion also creates a high conductivity artefact close to the surface. In addition, the interface between the two different layers is not identified. Figs \ref{fig:different_wavelets}(b) to \ref{fig:different_wavelets}(d) show inversion results for wavelets with increasing vanishing moment (and increasing regularity). To facilitate result comparison, the regularization parameter was fixed to $10^{-5}$ for all inversions (The L-curve criterion selected very similar regularization parameters for each wavelet). As expected from the shape of the wavelet, the db2-wavelet is capable of both showing sharp and relatively smooth transitions. For other wavelets, the inversion results are relatively identical and smoother, but with a larger peak than commonly found with Tikhonov regularization. So far, we could not determine a method to select an optimal wavelet from the data, as this remains strongly dependent on the expected geological variations. On the other hand, one can impose the characteristics of the inversion scheme as desired. For blocky inversion, one chooses a low number of vanishing moments, while for smooth inversion, a larger vanishing moment can be opted for. Note that in general the maximal electrical conductivity value is recovered more exactly with wavelet inversion than for smooth Tikhonov inversion (i.e. large number of vanishing moments) and that the results are more symmetrical near the actual conductive layer. The only disadvantage is the extra structure at larger depths. Those are oscillations which are a result of the shape of the wavelet (the more vanishing moments, the more oscillations as can be seen from Fig \ref{fig:scaling_functions} in Appendix \ref{appendix:daubechies}).\\

The scale-dependent wavelet based inversion is advantageous for recovering smoother inversion as well. As in the previous section, we compare the inversion with the conductivity profile from the alluvial aquifer. For each inversion, the L-curve criterion determines the optimal regularization parameter which is different for each wavelet. The results for Tikhonov, db1 and db6 wavelet inversion are shown in Fig \ref{fig:modelspacedb6}. The results for db1 and Tikhonov are similar. Both schemes recover peaks at the same locations with more or less the same values for electrical conductivity. As with the previous test with this profile, the wavelets with more vanishing moments can recover the peaks better. The case with the correct location of the peak in the absence of experimental noise is shown in Fig \ref{fig:nonoise2} in Appendix \ref{appendix:nonoise}.\\

The shape of the wavelet (and scaling function) affects the inversion result, and choosing the wavelet should be based on prior knowledge. This can be understood from the theoretical framework, namely by penalizing large coefficients of the wavelet representation of a model, the outcome is altered with exactly the shape of the wavelet or scaling function. Therefore, the shape is often recognized in the inversion result. Consider another example with a softer interface between two different layers. In Fig \ref{fig:westhoek}, such a conductivity profile from \citep{hermans2012imaging} is shown from De Panne, Flanders, where a saltwater lens is lying on a clay layer. We show the inversion results for 4 different wavelets: db1, db2, db3 and db6. In general, better results are obtained when more model parameters are considered for more vanishing moments. For each case, the Tikhonov smoothing inversion is also shown in the corresponding number of model parameters. For all those cases, the Tikhonov regularization cannot fit the actual peak, while the wavelet inversion can, independently from the chosen wavelet. We again observe a decreasing blockyness for larger vanishing moments and an artificial second peak for db3 and db6, due to the shape of the wavelet. Knowing about this feature, one can more easily identify the effect of regularization on the inversion results and combine the results of db1 and db6 can be combined to successfully interpret the data. 

\subsection{The role of the forward model operator}
\label{sec:role_F}
In the previous section in the absence of unmodelled errors $(||\mitbf{\kappa}|| = 0)$, our proposed inversion scheme successfully recovered blocky structures. Next, we test the effectiveness of our scheme with an approximated forward operator $\mathcal{F}_{\mathrm{approx}}$ instead of the exact computation $\mathcal{F}$, for the purpose of computational efficiency. The synthetic data is generated from the operator $\mathcal{F}$ to mimic a true survey (including 5\% multiplicative Gaussian noise), hence we examine our scheme in the presence of biased error $\mitbf{\kappa}$. Using the scheme for both the LIN approximation and the damped model allows comparing the effect of the degree of unmodelled errors on the result of the inversion, especially in a conductive setting where the LIN assumption breaks down.\\

\begin{figure}
	\centering
	\includegraphics[width= 0.495\textwidth]{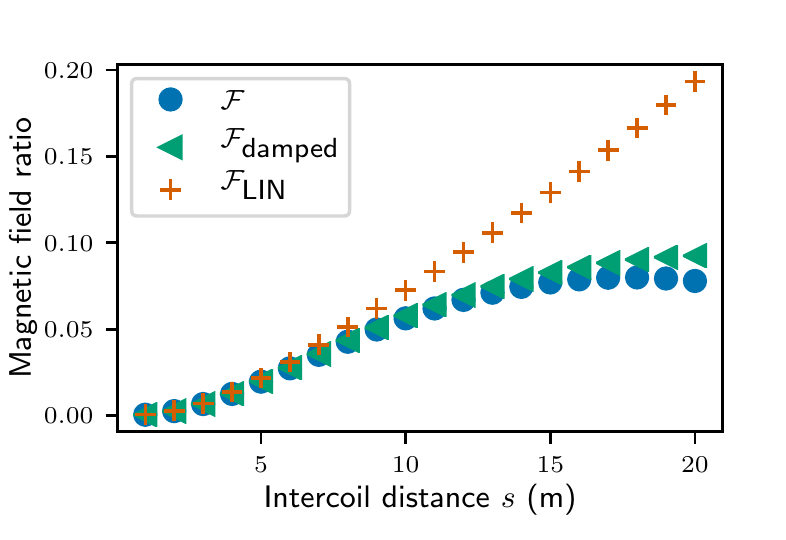}
	\caption{ Magnetic field ratio generated via the different forward operators from the conductivity profile in Fig \ref{fig:fw}.}
	\label{fig:dataplots}
\end{figure}

The magnetic field data generated from the exact model $\mathcal{F}$, the LIN approximation $\mathcal{F}_{\mathrm{LIN}}$ and the damped model $\mathcal{F}_{\mathrm{damped}}$ are shown in Fig \ref{fig:dataplots}. From this data, it is clear that the noise is biased and larger for the LIN approximation than for the damped model, i.e. their error is $||\mitbf{\kappa}_{\mathrm{LIN}}|| = 0.23$ and $||\mitbf{\kappa}_{\mathrm{damped}}|| = 0.026$ respectively. As expected, we get $||\mitbf{\kappa}_{\mathrm{damped}}|| < ||\mitbf{\kappa}_{\mathrm{LIN}}||$. In terms of data-misfit \eqref{eq:datamisfit}, the unbiased experimental noise equals 0.000149, the biased unmodelled error in damped and experimental noise equals 0.0002664, while for the LIN approximation this equals 0.00996. Note that the error and its norm will differ for different conductivity profiles.\\

\begin{figure*}
	\centering
	\includegraphics[width= 0.495\textwidth]{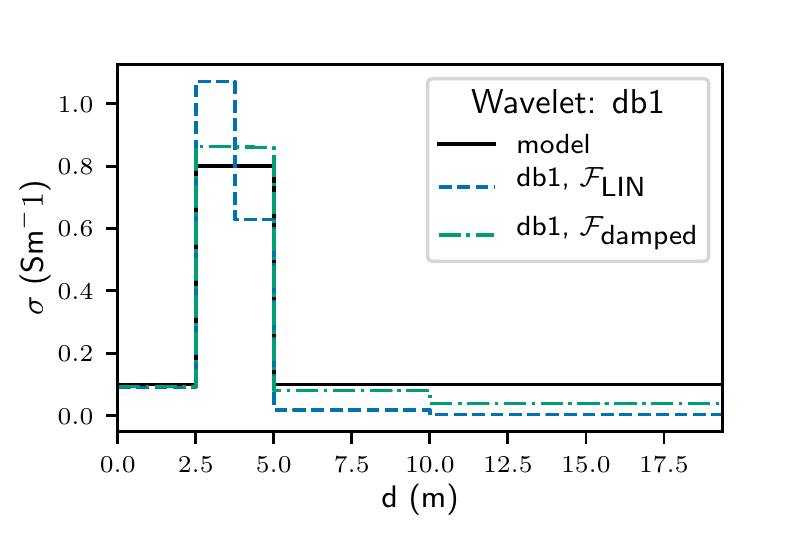}
	\includegraphics[width= 0.495\textwidth]{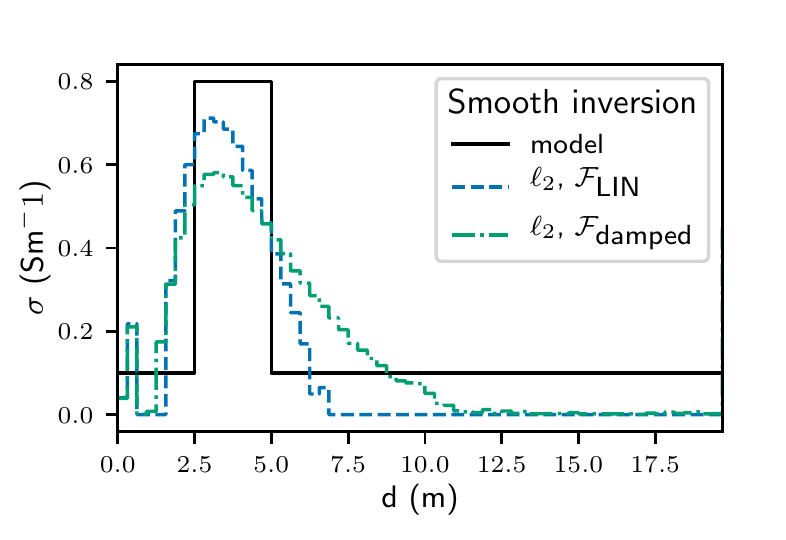}
	\caption{Effect of forward operator (and modelling error $\mitbf{\kappa}$) on the inversion result.}
	\label{fig:fw}
\end{figure*}

The L-curve criterion can easily locate vertical branch due to the experimental random noise, but we can not use this principle to estimate a regularization parameter such that the data misfit meets the value of the sum of the experimental and biased noise. Furthermore, we cannot rely on the discrepancy principle, since the biased noise $\mitbf{\kappa}$ is unknown. The optimal estimated regularization parameters following the L-curve criterion yield the estimates in the range of the experimental noise. Their data misfits $\phi_d$ for the LIN and damped model are 0.000217 and 0.000143 respectively. The damped model has a misfit approximately equal to the experimental noise level, while for LIN this is slightly larger. The results are shown in Fig \ref{fig:fw}. The conductivity profile obtained via the damped model has only a slightly deviant electrical conductivity, while for the LIN approximation slightly too much structure was introduced. However, both models recover the profile relatively well. Surprisingly, the smooth inversion with the LIN approximation yields better results. This example illustrates the importance of a good approximation for wavelet-based inversion, yet, our method seems relatively robust.\\

\subsection{Field data case}
\begin{figure}
	\centering
	\includegraphics{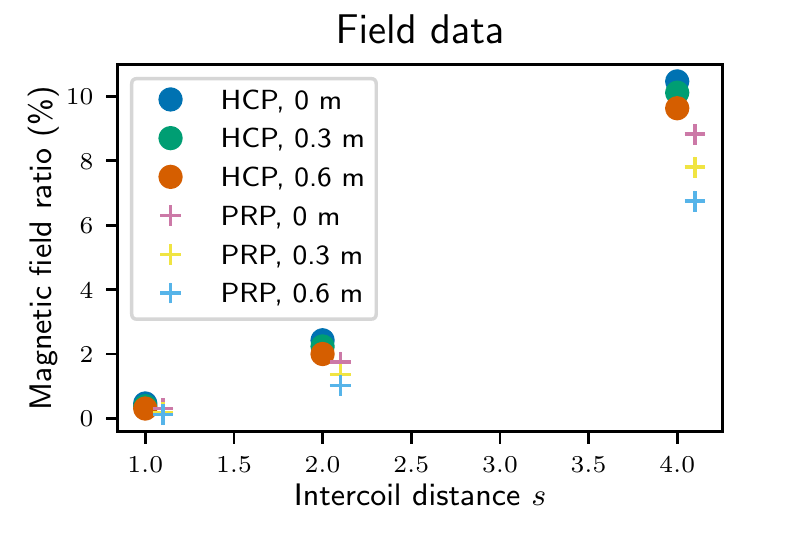}
	\caption{FDEM field data obtained with the DUALEM 421S, from \citet{bobe2020efficient}.}
	\label{fig:fielddata}
\end{figure}

In this section, we demonstrate the wavelet-based inversion scheme on real FDEM field data, obtained from \citep{bobe2020efficient}. The measurement location is in Gontrode forest, Belgium, known for its horizontal stratigraphy of different soil layers (i.e. no 2D or 3D effects are expected).  For the sounding, the DUALEM 421S was used with six receiver coils (three HCP coils at 1,2 and 4 metres and three PRP coils at 1.1,2.1 and 4.1 metres) with an operating frequency of 9.0 kHz. The sounding was performed at zero, 30 cm and 60 cm height. The magnetic field ratios are shown in Fig \ref{fig:fielddata}.\\

\begin{figure*}
	\centering
	\includegraphics[width= 0.48\textwidth]{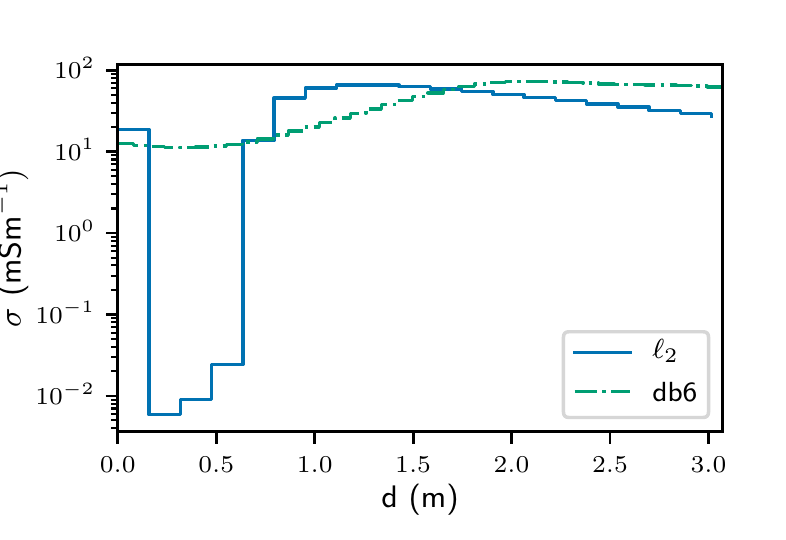}
	\includegraphics[width= 0.48\textwidth]{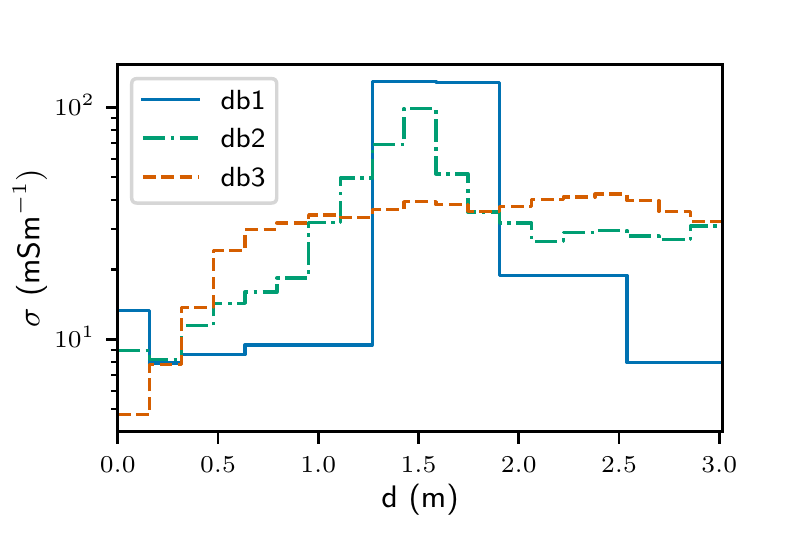}
	\caption{Inversion results from Tikhonov regularization ($\ell_2$) and wavelet-based regularization with Daubechies (db) wavelets. Estimation of regularization parameter via L-curve criterion.}
	\label{fig:fieldresults}
\end{figure*}

\citet{bobe2020efficient} conducted a visual inspection of the subsurface for the validation of the results, by drilling a hole with an Edelman auger. The top 0.15 meters consists of organic rich forest material with many air-filled voids. Between 0.15 and 0.8 metres, a clay-rich sand was observed. The clay content decreased with further depth. At 1.5 metres, the groundwater table was reached.\\

The result of the inversion procedure with the damped model as forward operator is shown for db1,db2,db3 and db6 wavelets and for smooth inversion in Fig \ref{fig:fieldresults}. The results of the inversion with other wavelets exhibited similar characteristics. We have used an equidistant parameterization of 128 model parameters up to 10 m depth and only the outcome of the first three meters is shown (because of the little sensitivity in the lower depth region). The results can be compared to the outcome in figure 7 (red line) in \citet{bobe2020efficient}, where with FDEM data alone, the clay-rich sand was not recovered. It shows a steady increase in electrical conductivity from 10 mS/m at the surface to almost 100 mS/m at two meters depth, followed by a constant plateau.\\

We expect the first 0.15 metres to be highly resistive (air voids are an insulator). Clay is known to have a relatively high bulk electrical conductivity value, which explains the peak in the conductivity profiles. Sand, on the other hand, has a low bulk electrical conductivity. Groundwater contains more free electrical charges and therefore, the electrical conductivity usually increases in the saturated zone. This behaviour is recovered in almost all wavelet-based inversion results, however a peak as observed in db1 and db2 is not  necessarily expected. Based on the lithological profile, db3 seems to provide the more realistic results. The main difference with smooth inversion from \citet{bobe2020efficient} is the increase in electrical conductivity due to the water table, which is not present in the smooth inversion and inversion with db6 and the unrealistic presence of the low conductivity between 15 and 50 cm depth. . \\

The wavelet-based regularization scheme yields results with similar characteristics, yet the results are different. This is the faith of deterministic inversion, however the wavelet-based inversion scheme would also suit in stochastic inversion methods.

\section{Conclusion}
We have introduced an improved inversion scheme for EMI surveys that can be extended to any other 1D geophysical method. It involves a new model misfit or regularization term based on the wavelet transform and scale-dependent weighting which can easily be combined with the existing framework of deterministic inversion (gradient-based optimization methods, L-curve criterion for optimal regularization parameter). \\

Our results show that the scale-dependent wavelet-based inversion scheme with sparsity constraint is more adaptive than Tikhonov regularization since it can recover both blocky and smooth conductivity profiles, by adequately choosing the number of vanishing moments of the wavelet. Furthermore, our scheme can recover high amplitude anomalies in combination with globally smooth profiles what is generally not possible, as discussed in Section \ref{sec:modwav}. The scale-dependency of our scheme allows to use of wavelets with few vanishing moments and is, therefore, an improvement with respect to existing wavelet-based regularization schemes. The adaptive nature of the inversion method allows for high flexibility because the shape of the wavelet can be exploited to generate multiple representations of the inverse model. Depending on available prior knowledge, the final result can be chosen or the results can be interpreted together. Although the choice os the wavelet is case-specific, we have observed few variations in the recovered model above db6. Together with the computational speed-up, our method thus offers a viable alternative to the existing methodologies, as discussed in the Introduction.\\

We have shown that the inversion scheme was robust even for approximate forward models. However, this should be confirmed in future work for more challenging conditions such as saline grounds and where a potential multimodal character of the objective function poses new challenges.

\begin{acknowledgments}
The authors thank Christin Bobe for making available the data from the field work in Gontrode, Belgium and Marieke Paepen for the conversations on the inversion of those data. The research leading to these results have gratefully received funding from FWO (Fund for Scientific Research, Flanders, grant 1113020N).
\end{acknowledgments}

\bibliographystyle{gji}

\appendix
\section{Daubechies Wavelets}
\label{appendix:daubechies}
\begin{figure*}
	\centering
	\includegraphics[width=1\textwidth]{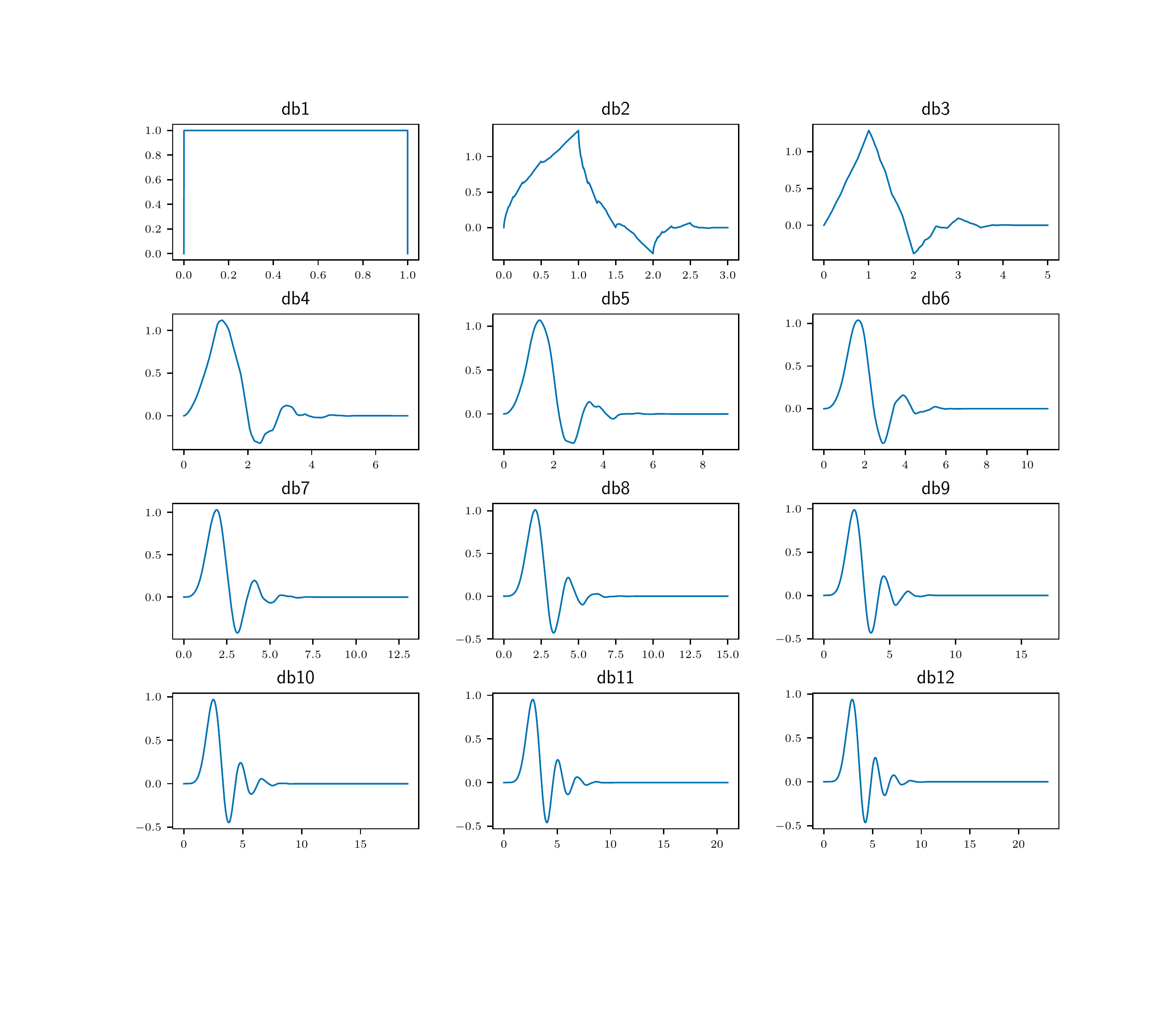}
	\caption{Scaling functions of the wavelets used in this paper.}
	\label{fig:scaling_functions}
\end{figure*}

Scaling functions of the wavelets used in this paper in Fig \ref{fig:scaling_functions}

\section{Results without experimental noise}
\label{appendix:nonoise}
\begin{figure}
	\centering
	\includegraphics[width=0.495\textwidth]{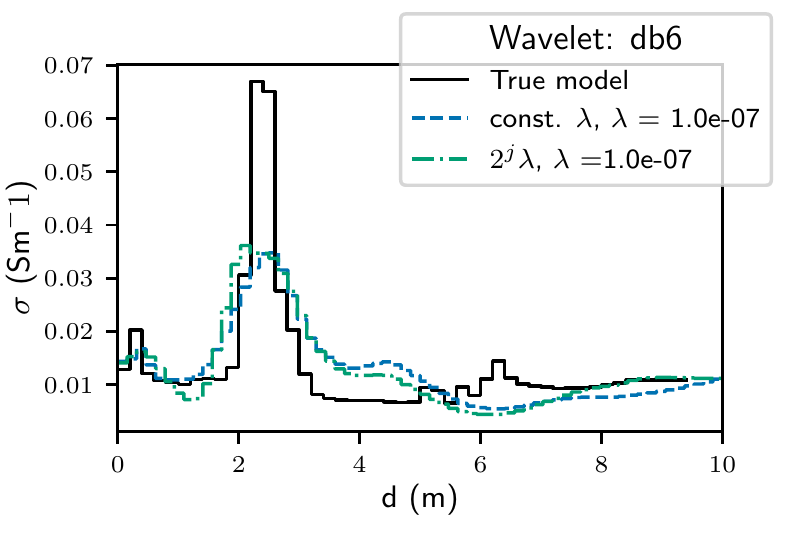}
	\caption{Identical inverse problem as in Section \ref{sec:scale_nonscale} but without any multiplicative noise.}
	\label{fig:nonoise1}
\end{figure}

\begin{figure}
	\centering
	\includegraphics[width=0.48\textwidth]{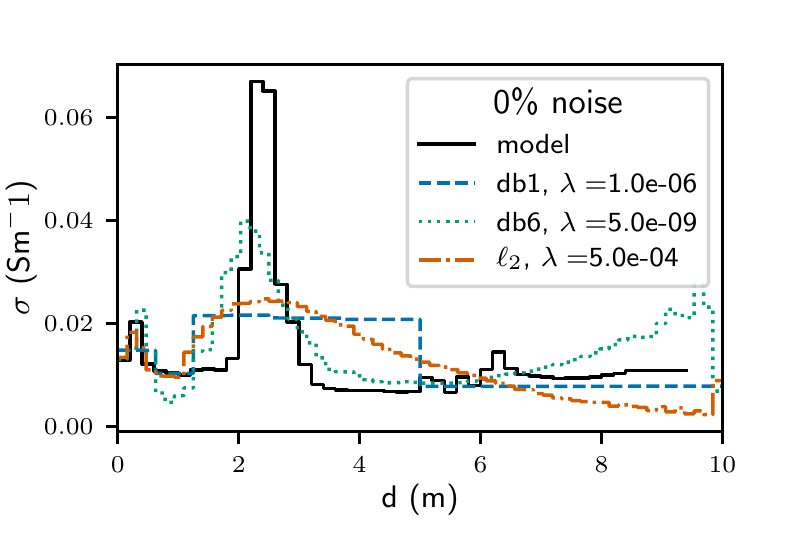}
	\caption{Identical inverse problem as in Section \ref{sec:modwav} but without any multiplicative noise.}
	\label{fig:nonoise2}
\end{figure}

Identical inverse problem as in Section \ref{sec:scale_nonscale} and \ref{sec:modwav} but without any multiplicative noise in Figs \ref{fig:nonoise1} and \ref{fig:nonoise2} respectively.

\bsp 

\label{lastpage}

\end{document}